\documentclass[twocolumn,epsfig,pre]{revtex4}
\usepackage{graphicx}
\usepackage{amssymb}
\usepackage{amsmath}
\usepackage{hyperref}
\usepackage{latexsym}

\def\be{\begin{equation}}
\def\ee{\end{equation}}
\def\bea{\begin{eqnarray}}
\def\eea{\end{eqnarray}}

\draft

\begin{document}

\title{Driven Transport on open filaments with inter-filament switching processes}
\author{Subhadip Ghosh$^{1}$, Ignacio Pagonabarraga$^{2}$ and Sudipto Muhuri$^{1,3}$}
\affiliation
{$^1$Institute of Physics, Sachivalaya Marg, Bhubaneswar 751005, India\\
$^2$ Departament de Fisica Fonamental, Universitat de Barcelona, C.Marti i Franques 1, 08028 Barcelona, Spain\\
$^3$ Department of Physics, Savitribai Phule Pune University, Ganeshkhind, Pune 411007, India}
\begin{abstract} 
We study a two filament driven lattice gas model with oppositely directed species of particles moving on two parallel filaments with filament switching processes and particle inflow and outflow at filament ends. The filament switching  process is {\it correlated} such that particles switch filaments with finite probability only when oppositely directed particles meet on the same filament. This model mimics some of the coarse grained features observed in context of microtubule (MT) based intracellular transport, wherein cellular cargo loaded and off-loaded at filament ends are transported on multiple parallel microtubule (MT) filaments and can switch between the parallel microtubule filaments. We focus on a regime where the filaments are weakly coupled, such that filament switching rates scale inversely as the length of the filament. We find that the interplay (off)loading processes at the boundaries and the filament switching process leads to some distinctive features of the system. These features includes occurrence of variety of phases in the system with inhomogeneous density profiles including localized density shocks, density difference across the filaments and bidirectional current flows in the system. We analyze the system by developing a mean field (MF) theory and comparing the results obtained from the MF theory with the Monte Carlo (MC) simulations of the dynamics of the system. We find that the steady state density and current profiles of particles and the phase diagram obtained within the MF picture matches quite well with  MC simulation results. These findings maybe useful for studying multi-filament intracellular transport. 
\end{abstract}

\maketitle
\section{Introduction}
One-dimensional driven diffusive systems, unlike their equilibrium counterparts are known to exhibit boundary induced phase transitions \cite{evans1,evans2,schutz}. Such systems have also served the purpose of providing a framework for studying wide class of driven biological phenomenon ranging from transport across biomembranes \cite{choubio}, to transport on individual cellular filament \cite{menon,freyprl,freypre,ignapre,madanpre} and cytoskeletal filament network \cite{kern}.

Filament based intracellular transport involves oppositely directed motors, which use multiple arrays of cytoskeletal filaments, to actively transport cellular cargoes such as mitochondria, endosomes, and pigment granules~\cite{cell,howard}. It has been observed that long-distance  cellular cargo transport on microtubule (MT) filaments is achieved by sets of oppositely directed motor proteins, e.g; {\it dynein} and {\it kinesin}, which attach to the cellular cargoes and  transport them {\it actively} along these filaments. The transport itself is determined by different processes at play at the molecular level, e.g; the motor processivity, directional switching dynamics of the cargo carried by the motors, the underlying filament organization, the (un)binding characteristics of the motors to the filament and the boundary input(output) rate of cargoes at filament ends~\cite{welte}. It has also been observed that both long distance regulated transport \cite{welte} and phenomenon of jamming arise out of the collective action of the these motor proteins~\cite{jamguna,jamleduc}. 

One of the approaches to study  transport  in such systems has been to describe it in terms of coarse-grained driven lattice gas models wherein the MT filament is considered as a 1-d lattice, the interactions between the transported cargoes are included via excluded volume effect and the underlying driven stochastic dynamics due to various processes incorporated in the description~\cite{ignapre}. Some of the previous theoretical attempts have focused on the interplay of stochastic directional switching mechanisms, directional hopping of individual cargoes and the effect of the input and output of the cargoes on the boundaries on a {\it single} filament~\cite{ignaepl,ignapre,madanpre}. However for a variety of biological situations such as axonal transport in neurons, cargo transport takes place on a  {\it multiple} parallel array of MT filaments~\cite{hollenbeck,jamleduc}.  For example, {\it in vitro} studies on cultured neurons have revealed that cargo switching between neighbouring filaments occurs in  axonal transport of mitochondria on neurons \cite{ross}. On theoretical grounds, it has been argued that even without considering the effects of the boundaries, the interplay of the translation process on filaments and the filament switching processes can manifest in form of a phase transition between a an inhomogeneous {\it jammed} phase of the transported cargoes and a freely flowing phase with homogeneous cargo  density  in each filament~\cite{jamepl}. Thus studying the role of multiple filaments in determining the transport properties of such systems is of considerable importance. 

Driven transport on parallel lattices have been studied theoretically in different contexts~\cite{pronina1,pronina2, harris, mitsudo, popkov1,popkov2, juhasz1, juhasz2, rolland, frey2lane, santen2,sugden,ignajstat,jamepl}, and the particle switching dynamics  between adjacent lanes have also been taken into account explicitly in some cases~\cite{jamepl,ignajstat,juhasz1,juhasz2, santen2, pronina1, pronina2,frey2lane,sugden}. In this paper we will focus on how the transport along  two parallel filaments is affected by the interplay of boundary inflow and outflow  of particles at the filament ends and filament switching dynamics of particles. Before we proceed  describing the mode in detail, we wish to highlight a few aspects of transport that have been observed in the context of intracellular transport : (a) Experimental studies, such as the one on endosomal transport on MT reveal that cellular cargoes can switch between neighbouring filaments~\cite{ross,roop}. (b) Experiments  suggest that cellular cargoes traveling in opposite direction on the same MT can also cross each other and continue with their translational motion along the same filament~\cite{roop}. (c) Cargo transport is also dependent on loading and offloading of cellular cargoes at the filament ends~\cite{welte,offload}.
 Motivated with these experimental observations, in the model that we study, we focus on the interplay of the boundary driven processes of particle input and output at filament ends with the {\it active} motor driven cargo translocation on the filaments and filament switching processes. Accordingly,  we will consider two parallel filaments with oppositely directed species of particles which translate on the lattices with a specified hopping rate. The oppositely directed particles are also allowed to  pass through each other with a certain specified rate on the same lattice. The particles are also allowed to switch between the lattices with certain finite probability {\it only } when oppositely directed species meet each other on the same lattice, so that the switching between the lane is a correlated process~\cite{ignajstat,jamepl}. Thus implicitly we will consider that  individual motors carrying the cargo have a propensity to switch between different filaments when they experience a force, when hindered by an oppositely directed particle moving on the same filament. This in turn can decrease the binding affinity of the motor to the filament and induce it to switch and bind to the neighbouring filament. Finally we will allow particles to enter and exit the filament ends with prescribed  rates.

In Section~\ref{sec:model} we specify the model, the dynamical rules on the lattice and the corresponding equations of motion for the system. 
In Section~\ref{sec:mf} we set up the Mean Field (MF) continuum equations for the system and analyze the boundary conditions at the ends of the lattice. In Section~\ref{sec:phase} we analyze the different phases possible for the system, obtain the corresponding MF steady states density profiles for the particles and construct the MF phase diagram for the system and compare these results with the Monte Carlo (MC) simulation results. Finally in Section~\ref{sec:conclusion} we discuss these results in context of multi-filament intracellular transport.

           
\section{The model}
\label{sec:model}

\begin{figure}[h]
\centering
\includegraphics[width=3.2in,angle=0]{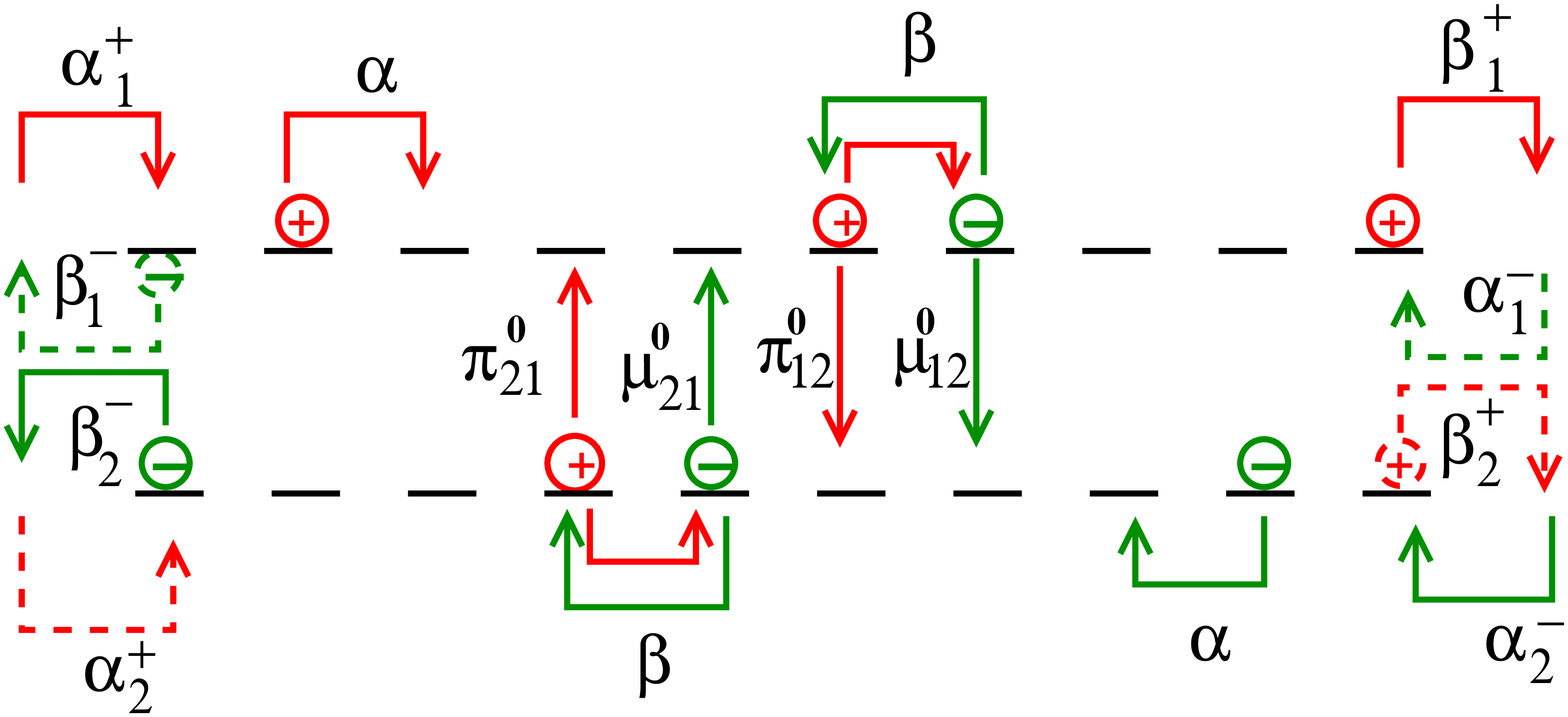}
\begin{center}
\caption{Schematic representation of the processes of translocation, switching between the two filaments, and entry/exit at filament boundaries.}
\end{center}
\end{figure}
We consider two parallel cellular filaments, represented by  two finite and parallel one dimensional lattices of length $L$ with $N$ sites, with  lattice spacing $\epsilon = L/N$. The cellular cargoes  transported along these filaments will be referred to as   particles, and will be characterized by two different species. Specifically, along each filament, the transported cargo can either be a $(+)$ particle which moves from left to right on the filament or a $(-)$ particle that moves from right to left. Without loss of generality, the  two  filaments are labeled as $1$ and $2$. The instantaneous state of the system is described in terms of the occupation numbers, which indicate the spatial localization of the two species of particles on the two parallel filaments. Specifically, $n_{i,1}^{+}$ corresponds to a occupation number of a particle at site $i$  moving to the $right$ on filament $1$. The maximum allowed occupancy at any lattice site is 1 so that each lattice site is occupied either by a $(+)$ particle, $(-)$ particle or is vacant, $(0)$.  

The dynamics of this system can be expressed in terms of the movements allowed for the particles. For the sites in the {\it bulk}, in each individual filament, a $(+)$ particle can hop from site $i$ to site $i+1$ with a rate $\alpha$ if that site is vacant. Similarly a $(-)$ particle can hop from site $i$ to site $i-1$ with an identical rate $\alpha$ if it is vacant. If the $i$-th site on filament 1 is occupied by a $(+)$ and if the neighboring site to the right, i.e; site $i+1$ is occupied by an oppositely directed $(-)$, then two different  processes can occur:  with rate  $\beta$ the two particles can swap their positions, and  the $(+)$ particle  moves to the site $i+1$ and the oppositely moving $(-)$ particle  moves to site $i$; while  with rate $\pi_{12}^{o}$ the  $(+)$ particle can switch the to filament-$2$, at the same corresponding  site, identified by its  index $i$,  if the site  is vacant.
Similarly, a $(-)$ particle from filament $1$ at site $i+1$ can switch to the site of filament $2$ with the same site index $i+1$ with a rate $\mu_{12}^{o}$ if that site on the other filament is vacant. Identical processes that we have described for filament $1$ also happens for filament $2$. The rates of filament switching processes from filament $2$ to filament $1$ for the $(+)$ and $(-)$ particles are $\pi_{21}^{o}$ and  $\mu_{21}^{o}$ for $(+)$ and $(-)$ particles, respectively. Particle switching  between the two filaments can be understood as arising from the stronger loading force experienced by the motor proteins which carry the cargo when they push against an  oppositely directed cargo. This leads to an increase in the rate of motor detachment  (along with the cargo) from the filament and offers the possibility of a subsequent reattachment of the molecular motor to the neighbouring site on the  other filament~\cite{lipo}. Since we are interested in the regime where the filaments are {\it weakly} coupled, we probe the regime where the filament switching processes compete with loading and offloading processes at filament boundaries. We systematically implement it by choosing the filament switching rates at individual lattice site such that they scale inversely with  system size so that we have, $\pi_{12}^{o} = \frac{\pi_{12}}{N}$, $\mu_{12}^{o} = \frac{\mu_{12}}{N}$, $\pi_{21}^{o} = \frac{\pi_{21}}{N}$ and $\mu_{21}^{o} = \frac{\mu_{21}}{N}$. 

Although the bulk processes are analogous to those introduced in Ref.~\cite{ignajstat,jamepl},  which focused on  the collective behavior of cargoes moving along  filaments with a {\it closed} ring morphology and with overall particle number conservation, we will concentrate on the behavior of such particles for open filaments. 
In this configuration the overall particle number is not conserved, and the motion on incoming and outgoing particles at the filament ends must be accounted for.

For the {\it boundary sites} at the filament ends, a $(+)$ particle can enter the left end of the filament (with site label $i = 0$) of filament $1$ with a rate $\alpha_{1}^{+}$ if it is vacant, and it can leave from the right end of the filament (with site label $N-1$) of filament $1$ with a rate $\beta_{1}^{+}$. Similarly a $(-)$ particle can enter the right end of the filament $1$ with a rate $\alpha_{1}^{-}$ if it is vacant and it can leave from the left end of the filament with a rate $\beta_{1}^{-}$. Similar processes also occur in filament $2$ with the corresponding rates being $\alpha_{2}^{+}$, $\beta_{2}^{+}$, $\alpha_{2}^{-}$ and $\beta_{2}^{-}$ respectively. All the the dynamic processes that characterize the model are schematically depicted in Fig.~1. 
\section{Mean Field Evolution Equations}
\label{sec:mf}
The time evolution for the average occupation number for the two oppositely directed species along each individual filament can be expressed in terms of gain and loss terms arising from translation and filament switching processes. As described in Ref.~\cite{ignajstat}, these terms involve the averages of the local occupation numbers for the particles at each site as well as various combination of averages of two-point correlators of site occupation numbers, which account for the role of particle correlations in the particle collective dynamics.

We set  $\alpha = \beta = 1$ and  choose $\pi_{12} = \pi_{21} = \mu_{12} = \mu_{21} = \pi$, which correspond to a symmetric scenario  where the propensity to switch filaments is the same for both $(+)$ and $(-)$ species and it is symmetric about filament $1$ and filament $2$. The Mean Field (MF) evolution equations are obtained  factorizing the two point correlators of the occupation numbers. The continuum  MF evolution equations are derived by rescaling the total length $L$ to $1$ and letting $N \rightarrow \infty$ so that $\epsilon \rightarrow 0$~\cite{freyprl,ignajstat}.

Correspondingly,  $p_{1}(x)$, $p_{2}(x)$, $n_{1}(x)$ and $n_{2}(x)$ are then the  average densities as a function of the relative position in the filament, $x$. The MF continuum equations in the bulk can be expressed as,

\begin{eqnarray}
\partial_{t}p_{1}&=& \epsilon\pi \left[ p_{2}n_{2}( 1 - p_{1} - n_{1}) -  p_{1}n_{1}( 1 - p_{2} - n_{2}) \right] \nonumber\\
&-&\epsilon\partial_{x}\left[p_{1}( 1 - p_{1})\right] + O(\epsilon^{2})
\label{mf-eqn1}
\end{eqnarray}
\begin{eqnarray}
\partial_{t}p_{2}&=& \epsilon\pi \left[ p_{1}n_{1}(1 - p_{2} - n_{2}) - p_{2}n_{2}( 1 - p_{1} - n_{1}) \right] \nonumber\\
&-&\epsilon \partial_{x}\left[p_{2}( 1 - p_{2})\right] + O(\epsilon^{2})
\label{mf-eqn2}
\end{eqnarray}
\begin{eqnarray}
\partial_{t}n_{1}&=& \epsilon\pi \left[ p_{2}n_{2}( 1 - p_{1} - n_{1}) - p_{1}n_{1}( 1 - p_{2} - n_{2}) \right] \nonumber\\
&+&\epsilon \partial_{x}\left[ n_{1}( 1 - n_{1}) \right] + O(\epsilon^{2})
\label{mf-eqn3}
\end{eqnarray}
\begin{eqnarray}
\partial_{t}n_{2}&=& \epsilon\pi \left[ p_{1}n_{1}( 1 - p_{2} - n_{2}) - p_{2}n_{2}( 1 - p_{1} - n_{1}) \right] \nonumber\\
&+&\epsilon \partial_{x}\left[n_{2}( 1 - n_{2})\right] + O(\epsilon^{2})
\label{mf-eqn4}
\end{eqnarray}
where we have displayed terms up to first order in  $\epsilon$. The corresponding expression for the currents of each species read, 
\begin{eqnarray}
J_{1}^{+} &=& p_{1}(1 - p_{1})\\
J_{1}^{-} &=& -n_{1}(1 - n_{1})\\
J_{2}^{+} &=& p_{2}(1 - p_{2})\\
J_{2}^{-} &=& -n_{2}(1 - n_{2})
\end{eqnarray}
\subsection{Steady State profiles}
\label{sec:ss}
From Eqs.(\ref{mf-eqn1})-(\ref{mf-eqn4}) , the steady state profiles corresponding to the continuum MF evolution of the molecular motors satisfy, 
\begin{eqnarray}
\frac{d J^{+}_{1}}{d x} = \pi [p_{2}n_{2}(1 - p_{1} - n_{1}) - p_{1}n_{1}(1 - p_{2} - n_{2})]
\label{eq:evo+1}\\
\frac{d J^{-}_{1}}{d x} = \pi [p_{2}n_{2}(1 - p_{1} - n_{1}) - p_{1}n_{1}(1 - p_{2} - n_{2})]
\label{eq:evo-1}\\
\frac{d J^{+}_{2}}{d x} = \pi [p_{1}n_{1}(1 - p_{2} - n_{2}) - p_{2}n_{2}(1 - p_{1} - n_{1})]
\label{eq:evo+2}\\
\frac{d J^{-}_{2}}{d x} = \pi [p_{1}n_{1}(1 - p_{2} - n_{2}) - p_{2}n_{2}(1 - p_{1} - n_{1})],
\label{eq:evo-2}
\end{eqnarray}
which govern the bulk profiles of $(+)$ and $(-)$ particles in the  two filaments. One can rewrite them in terms of the  fluxes of the total number of particles and its difference introducing 
\begin{eqnarray}
J_{p} = p_1(1 - p_1) + p_2(1 - p_2)
\label{eq:cur5} \\
J_{n} = n_1(1 - n_1) + n_2(1 - n_2)
\label{eq:cur6}
\end{eqnarray}
Since in the bulk the total current for the combined system comprising of the two filaments is  isolated,  $J_{p}$ and $J_{n}$ have to be  spatially constant.
Subtracting Eq.~(\ref{eq:evo-1}) from Eq.~(\ref{eq:evo+1}) we  identify an additional conservation law,
\begin{eqnarray}
J_{1} = p_1(1 - p_1) + n_1(1 - n_1)
\label{eq:cur3}
\end{eqnarray}
which corresponds to the sum of the absolute values of the currents of opposite species along one filament. As shown in Appendix B, it is useful to reorganize the three independent conserved quantities, $J_{p}$, $J_{n}$ and $J_{1}$ and which remain spatially uniform in the bulk, in terms of   three new parameters, $J_{2} = J_{p} - J_{1} + J_{n}$, $C_{1} = J_{p} -J_{2}$ and $C_{2} = J_{p} - J_{1}$. Using these new conserved quantities, the equations for the density profiles  in the bulk  can be decoupled, and express them in terms of one single density variable. Specifically, we can write
\begin{eqnarray}
\frac{dp_1}{dx} &=& -\frac{1}{1 - 2p_1}[p_1\eta_{p_1}^{\pm}\left(1 - \mu_{p_1}^{\pm} - \nu_{p_1}^{\pm}\right)\nonumber\\ 
&-& \mu_{p_1}^{\pm}\nu_{p_1}^{\pm}\left(1 - p_1 - \eta_{p_1}^{\pm}\right)]
\label{eq:decup+1}
\end{eqnarray}
where $\eta_{p_1}^{\pm},\mu_{p_1}^{\pm}$ and $\nu_{p_1}^{\pm}$ are functions of $p_{1}$ alone. Their explicit functional dependence is provided in Appendix B. Similarly, we can get decoupled differential equations for $n_1$, $p_{2}$ and $n_{2}$,
\begin{eqnarray}
\frac{d n_1}{dx} &=& \frac{1}{1 - 2n_1}[n_1\eta_{n_1}^{\pm}\left(1 - \mu_{n_1}^{\pm} - \nu_{n_1}^{\pm}\right)\nonumber\\ 
&-& \mu_{n_1}^{\pm}\nu_{n_1}^{\pm}\left(1 - n_1 - \eta_{n_1}^{\pm}\right)]
\label{eq:decup-1}
\end{eqnarray}
\begin{eqnarray}
\frac{dp_2}{dx} &=& -\frac{1}{1 - 2p_2}[p_2\eta_{p_2}^{\pm}\left(1 - \mu_{p_2}^{\pm} - \nu_{p_2}^{\pm}\right)\nonumber\\ 
&-& \mu_{p_2}^{\pm}\nu_{p_2}^{\pm}\left(1 - p_2 - \eta_{p_2}^{\pm}\right)]
\label{eq:decup+2}
\end{eqnarray}
\begin{eqnarray}
\frac{dn_{2}}{dx} &=& \frac{1}{1 - 2n_2}[n_1\eta_{n_2}^{\pm}\left(1 - \mu_{n_2}^{\pm} - \nu_{n_2}^{\pm}\right)\nonumber\\ 
&-& \mu_{n_2}^{\pm}\nu_{n_2}^{\pm}\left(1 - n_2 - \eta_{n_2}^{\pm}\right)]
\label{eq:decup-2}
\end{eqnarray}
The explicit form of the decoupled differential equations which govern the density profiles and the relevant solutions are discussed in Appendix B. In Appendix B, Eq.(\ref{eq:p1}-\ref{eq:n2}), provides the mathematical expression for the quantities present in Eq.(\ref{eq:decup+1}- \ref{eq:decup-2}). Appendix B also describes the different sets of density profiles that can be obtained from the previous set of equations. We have found that there are 16 different branch solutions corresponding to the decoupled differential equation. The choice of a particular boundary condition corresponding to a particular phase, restricts the possible  choices to 8. Finally as discussed in Appendix B, physical considerations such as bounds on the physical value of density selects an unique solution to these differential equations for each set of prescribed boundary conditions. 
Therefore, the continuous MF equations determine the density profiles of the particles in the two filaments, once the boundary conditions are prescribed.  Although Eqs.(\ref{eq:decup+1}- \ref{eq:decup-2}) for the  different species densities decouple in the bulk, they are coupled through the boundary condition of Eqs.(\ref{eq:cur5}- \ref{eq:cur3}).  In the next subsection we discuss the set of boundary conditions satisfied by these differential equation corresponding to a particular phase. 

\subsection{Boundary conditions}
\label{sec:bc}
The allowed phases that characterize the   state of transport on the two filaments  is controlled by the particle input and output at the boundaries. The steady state density profiles are determined by  first order differential equations. This is due to the fact that the diffusive contribution is of higher order in the lattice spacing, $\epsilon$, and their contribution drops in the continuum limit, $\epsilon\rightarrow 0$.  As a result, the system cannot fulfill, generically, the input and output boundary conditions and   boundary layers are expected~\cite{freyprl}. One then must determine which of the fluxes at the filaments' ends determine the bulk density profiles and under which conditions coexistence between density phases can develop along the filaments. 

For the system composed of two filaments, we have 8 different particle entrance or exit rates at the boundaries, that we express as  $\alpha^{\pm}_{1/2}$,$\beta^{\pm}_{1/2}$. However there are only 4 boundary conditions to be specified for the complete solution of the steady state differential equations. In order to figure out the possible physically relevant boundary conditions, it is useful to build upon the boundary conditions that are satisfied by a closely related 1-D lattice gas model~\cite{evans1,evans2}, which has the similar translocation dynamics along the filament as our model, but which does not allow for  inter-filament exchange processes. In fact the model discussed in Ref.~\cite{evans1,evans2} is exactly the same as ours for the particular case of $\pi=0$, which corresponds to a situation where the inter-filament switching dynamics of the particles is turned off. Since for our case, the filaments are weakly coupled thus it is expected that the boundary conditions satisfied for our two-filament system are the various possible combination of the boundary conditions that are satisfied for individual lattice for the case studied in Ref.~\cite{evans1,evans2}. However we would like to stress that although the boundary conditions are obtained as simple combination of boundary conditions of the individual lattices, the resultant density and current profile obtained by spatially integrating the steady state differential equation would be qualitatively different due to the lattice switching term in the bulk.

We enumerate the possible combination of the boundary conditions and the resultant phases for each of those particular combinations.

{\it Filament $1$ in $HL$ phase and filament $2$ in $HL$ phase} ($(HL)_{1}-(HL)_{2}$): When the bulk current of $(+)$ matches with the output current of $(+)$ at the right boundary and the input current of $(-)$ matches with the bulk current of $(-)$ at the right boundary both for filament $1$ and $2$, the resultant phase corresponds to a situation where the $(+)$ particles are in {\it high density}(H) phase and $(-)$ particles are in {\it low density}(L) phase (refered to as $HL$ phase), in both the filaments. The boundary conditions that are satisfied in the continuum limit are,
\begin{eqnarray}
J_{1R}^+ &=& \beta_1^+p_{1R} = p_{1R}(1 - p_{1R}) \nonumber\\
J_{1R}^- &=& -\alpha_1^-(1 - p_{1R} -n_{1R}) = -n_{1R}(1 - n_{1R})\nonumber\\
J_{2R}^+ &=& \beta_1^+p_{1R} = p_{1R}(1 - p_{1R}) \nonumber\\
J_{2R}^- &=& -\alpha_1^-(1 - p_{2R} -n_{2R}) = -n_{2R}(1 - n_{2R})
\label{eq:hlhl}
\end{eqnarray}
where $J_{1R}^+$,  $J_{1R}^-$,  $J_{2R}^+$ and  $J_{2R}^+$ refer to the currents for $(+)$ and $(-)$ particles  in filaments $1$ and $2$ at the right(R) boundary respectively,  while $p_{1R}$, $n_{1R}$, $p_{2R}$ and $n_{2R}$ refer to the densities of $(+)$ and $(-)$ particles in filaments $1$ and $2$ at the right boundary. 
Using Eq.(\ref{eq:hlhl}), we can find the expression of the boundary densities at the right end of both filaments in terms of the entry and exit  particle rate 
\begin{eqnarray}
p_{1R} &=& 1 - \beta_1^+\label{eq:HD1den+}\nonumber\\
n_{1R} &=& \frac{(1+\alpha_1^-) - \sqrt{(1+\alpha_1^-)^2 - 4\alpha_1^-\beta_1^+}}{2}\label{eq:LD1den-}\nonumber\\
p_{2R} &=& 1 - \beta_2^+\label{eq:HD2den+}\nonumber\\
n_{2R} &=& \frac{(1+\alpha_2^-) - \sqrt{(1+\alpha_2^-)^2 - 4\alpha_2^-\beta_2^+}}{2}\label{eq:LD2den-}
\end{eqnarray}
By symmetry  there can be another phase where both in filament 1 and 2, $(-)$  are in {\it high density} phase while the $(+)$ are in {\it low density} phase and the current conditions are satisfied at the left boundary. Further one can find a  situation where for filament 1 the current at the left boundary for $(+)$ and $(-)$ matches with the bulk current, while for filament 2, the current at the right boundary for (+) and (-) matches with the bulk current. Similarly, there exists a   phase for where  for filament 1, the current at the right boundary for $(+)$ and $(-)$ matches with the bulk current, while for filament 2, the current at the right boundary for $(+)$ and $(-)$ matches with the bulk current. For all these 4 different phases, the structure of the boundary condition is exactly similar.

{\it Filament $1$ in $LL$ phase and filament $2$ in $LL$ phase :} ($(LL)_{1}-(LL)_{2}$): When the bulk current of $(+)$ particle matches with the input current of $(+)$ at the left boundary and input current of $(-)$ matches with the bulk current of $(-)$ at the right boundary both for filament $1$ and $2$, the resultant phases corresponds to $LL$ phase in both the filaments. The boundary conditions satisfied by the currents are,

\begin{eqnarray}
J_{1L}^+ &=& \alpha_1^+(1 - p_{1L} -n_{1L}) = p_{1L}(1 - p_{1L})\nonumber \\
J_{1R}^- &=& -\alpha_1^-(1 - p_{1R} -n_{1R}) = -n_{1R}(1 - n_{1R})\nonumber\\
J_{2L}^+ &=& \alpha_2^+(1 - p_{2L} -n_{2L}) = p_{2L}(1 - p_{2L})\nonumber\\
J_{2R}^- &=& -\alpha_2^-(1 - p_{2R} -n_{2R}) = -n_{2R}(1 - n_{2R})
\label{eq:cond52}
\end{eqnarray}
while the  expression for the currents at the other boundaries read 
\begin{eqnarray}
J_{1R}^+ = \beta_1^+p_{1R}\nonumber\\
J_{1L}^- = \beta_1^-n_{1L}\nonumber\\
J_{2R}^+ = \beta_2^+p_{2R}\nonumber\\
J_{2L}^- = \beta_2^-n_{2L}\label{eq:cond42}
\end{eqnarray}
Using Eq.~(\ref{eq:cond52}) and Eq.~(\ref{eq:cond42}), we can express the boundary densities as a function of entry and exit rates and the currents~\cite{evans2}, 

\begin{eqnarray}
p_{2L} &=& \frac{\alpha_2^+(1 - p_{2L} -n_{2L})}{1 - p_{2L}} = \frac{J_{2L}^+}{J_{2L}^+/\alpha_2^+ + J_{2L}^-/\beta_2^-}\nonumber\\
p_{2R} &=& \frac{\beta_2^+p_{2R}}{1 - p_{2R}} = \frac{J_{2R}^+}{1 - J_{2R}^+/\beta_2^+}\nonumber\\
n_{2R} &=& \frac{\alpha_2^-(1 - p_{2R} -n_{2R})}{1 - n_{2R}} = \frac{J_{2R}^-}{J_{2R}^-/\alpha_2^- + J_{2R}^+/\beta_2^+}\nonumber\\
n_{2L} &=& \frac{\beta_2^-n_{2L}}{1 - n_{2L}} = \frac{J_{2L}^-}{1 - J_{2L}^-/\beta_2^-}\label{eq:cond72}
\end{eqnarray}
As opposed to the continuity of the overall particle fluxes at filament's ends due to particle conservation, $J_{pL} = J_{pR}$ and $J_{nL} = J_{nR}$ , the current on the left and right end of an individual filament track will in general differ due to particle filament switching. In order to determine the densities for the two types of particles in this phase, we will assume that  the $(+)$ current in left boundary of filament $1$  equals the  $(+)$ current in right boundary of the same filament, $\left( J_{1L}^+\right) =\left( J_{1R}^+\right)$. Similarly we use the same equality for  $(-)$ particles, $\left( J_{1R}^-\right)=\left( J_{1L}^-\right)$. Analogously,  we equate the currents at the left and the right boundary in filament $2$, $J_{1L}^{\pm} = J_{1R}^{\pm}$ and $J_{2L}^{\pm} = J_{2R}^{\pm}$. This is a reasonable assumption because tracks are weakly coupled, as has been checked with MC simulation.

This fact allows us to obtain an expression for the boundary densities in this phase. Explicitly, from Eq.(\ref{eq:cond52}) and Eq.(\ref{eq:cond72}), the boundary densities for both the lanes read , 
\begin{eqnarray}
p_{1L} = 1 - \frac{1}{\alpha_1^+}p_{1L}(1 - p_{1L}) - \frac{1}{\beta_1^-}n_{1R}(1 - n_{1R})\label{eq:ASLD-LDCON3}\nonumber\\
n_{1R} = 1 - \frac{1}{\beta_1^+}p_{1L}(1 - p_{1L}) - \frac{1}{\alpha_1^-}n_{1R}(1 - n_{1R})\label{eq:ASLD-LDCON4}\nonumber\\
p_{2L} = 1 - \frac{1}{\alpha_2^+}p_{2L}(1 - p_{2L}) - \frac{1}{\beta_2^-}n_{2R}(1 - n_{2R})\label{eq:ASLD-LDCON1}\nonumber\\
n_{2R} = 1 - \frac{1}{\beta_2^+}p_{2L}(1 - p_{2L}) - \frac{1}{\alpha_2^-}n_{2R}(1 - n_{2R})\label{eq:ASLD-LDCON2}
\end{eqnarray}
These coupled algebraic equations for each filament track can be numerically  solved to get the corresponding  boundary densities from which the density profiles can be numerically derived.

{\it Filament $1$ in $LL$ phase and filament $2$ in $HL$ phase :} ($(LL)_{1}-(HL)_{2}$): When for filament $1$, the bulk current of $(+)$ matches with the input current of $(+)$ at the left boundary and input current of $(-)$ matches with the bulk current of $(-)$ at the right boundary and the bulk current of $(+)$ matches  the output current of $(+)$ at the right boundary and the  input current of $(-)$ matches with the bulk current of $(-)$ at the right boundary in filament $2$, the resultant phase for the system corresponds to $LL$ phase in filament $1$ and $HL$ phase in filament $2$. 

In this case we again assume the continuity of the fluxes separately for the two particle types along filament 1, $J_{1L}^+ = J_{1R}^+$ and $J_{1L}^- = J_{1R}^-$. Accordingly, the equations for the boundary densities are similar in form to those expressed in Eq.~(\ref{eq:ASLD-LDCON2}) and the boundary densities for filament $1$ can be obtained numerically as discussed earlier. For filament $2$, the densities are determined by Eq.(\ref{eq:LD2den-}). Due to the symmetry in the swapping rates between the two filament, a second phase with analogous structure is feasible,where filament $1$ in $HL$ phase and filament $2$ in $LL$. There is still a further  symmetry associated with the HL phase in any of the two filaments, i.e; if the current of $(-)$ and $(+)$ at the left boundary matches with the currents in the bulk, the structure of the boundary conditions remains unaltered. In order to illustrate this, consider HL phase in a particular filament, then the corresponding densities are determined by  Eq.(\ref{eq:LD2den-}) and the boundary condition at $x =1$ is satisfied, so that  $(+)$  is in high density phase and $(-)$ is in low density phase. But analogously we could have a situation where the boundary condition at $x = 0$ is satisfied with $(+)$ particles in Low density phase and $(-)$ in high density phase. This situation would correspond to a different overall phase, but structure and the form of boundary density would be the same as Eq~.(\ref{eq:LD2den-}) with the only difference that the indexes of $(+)$ and $(-)$ in the expression for the boundary density in  Eq~.(\ref{eq:LD2den-}) is interchanged. Thus this structure of boundary conditions would correspond to 4 distinct phases. 

\section{Phases and phase diagram}
\label{sec:phase}

Since the boundary fluxes control the particle fluxes in the bulk,  once we have determined   the expression for the densities at the filament boundaries in terms of the input and output rates at the filament boundaries, the MF density and current profiles in the bulk can be determined from Eq.(\ref{eq:decup+1}-\ref{eq:decup-2}).  However, unlike the case of model in Ref.{\cite{evans1,evans2}} where the steady state density profiles and the corresponding phases in the bulk are solely determined by the boundary fluxes, the density profiles and phases now emerge from  the interplay of the boundary processes and particle filament  switching. This distinctly alters the nature of the density profiles and the topology of the phase diagram for the system under study. 

Specifically, a first major consequence of particle exchange between filaments is that  the density and current profiles in the bulk  are no longer spatially homogeneous.  In fact for certain range of input and output particle rates, the competition between the bulk and the boundary processes can result in density shocks along the filaments which are localized in the bulk. 

 Moreover, particle change between  filaments also allows for {\it phase coexistence} in the bulk apart from the {\it pure} phases which satisfies only one set of boundary conditions. This feature of phase coexistence occurs only because the current profiles in the bulk for a set of boundary conditions are not homogeneous so that different MF solutions for the current, intersect each other at specific spatial location in the bulk of the system. The phase coexistence in the bulk happens when the currents of the different MF solutions arising out of different boundary conditions match at a point in the bulk of the two-filament system. In that case part of the system obeys one set of the boundary conditions while the other half obeys another set of boundary condition and these set of solutions are joined by the condition of continuity of current at a particular location in the bulk of the lattice. In general the system selects the set of MF steady state solution for which the corresponding current is minimum. This holds true as long as the MF current profiles do not attain the maximal current value in the bulk. In this paper we have restricted our analysis to the region of parameter space of entry and exit rates of particles for which the condition for maximal current is not reached.

As a result of these new features, the topology of the resulting phase diagram  changes qualitatively with respect to the collective behavior in the absence of inter-filament particle exchange.  In the following subsection we first describe the procedure to find the MF density and current profiles and  determine the resultant phase. Subsequently we discuss the topology of the resultant phase diagram for this system and obtain the equations for the phase boundaries.    
 
\subsection{Density profiles and emerging phases}

In order to find the MF density and current profiles using Eq.(\ref{eq:decup+1}-\ref{eq:decup-2}), we have to first determine the three independent conserved currents in the system e.g; $J_{p}$, $J_{n}$ and $J_1$. Subsequently, we have to provide the appropriate values of the densities at the boundaries to completely specify the solutions for the individual species.


{\it $(HL)_1-(HL)_2$ phase}: Here, first we determine the values of the boundary densities at the right end of both filaments, e.g; $p_{1R}$, $n_{1R}$, $p_{2R}$ and $n_{2R}$ using Eq.(\ref{eq:HD1den+}). Thus $J_{p}$, $J_{n}$ and $J_{1}$ can be determined at the right boundary. The entire density profile can now be found out by evolving the MF solution from the left end of both filaments using Eq.(\ref{eq:decup+1}-\ref{eq:decup-2}). In Fig.2, we show the comparison of the MF profile with the MC simulations for this phase. It illustrates that both the density and the current profiles in the two filaments are not spatially homogeneous in contrast to similar phases in Ref.~{\cite{evans1,evans2}}. A similar procedure can be used to find out the profiles for the corresponding  $(LH)_1-(LH)_2$ phase.

{\it $(LL)_1-(HL)_2$ phase}: The values of the boundary densities, $p_{1L}$ and $n_{1R}$, can be determined by numerically solving Eq.(\ref{eq:ASLD-LDCON3}) for the first filament. For the second filament we use Eq.(\ref{eq:HD1den+}) to determine  $p_{2R}$ and $n_{2R}$. Thus, both $J_{n}$ and $J_{2}$ are identified at the right boundary. To determine $J_p$ we use the method of successive iterations. In the first iteration we set $p_{1R} = p_{1L}$ and obtain $J_p$ to get the density profiles. Again these profiles are not uniform due to inter-filament switching processes. Consequently, the density value obtained at $x = 0$ by evolving Eq.~(\ref{eq:decup+1}) from $x = 1$ is not same as $p_{1L}$. Hence, we take the difference between these two values at $x = 0$, and this difference is added to $p_{1R}$ and evolve it  to get the new density profile. We repeat this process until convergence is reached ~\footnote{ We have checked that  a relative  accuracy of $\sim 10^{-5}$ is enough to achieve significant results.}. This procedure allows then to derive  the entire density profile by evolving the MF solution from the right end of both filaments using Eq.(\ref{eq:decup+1}-\ref{eq:decup-2}). An analogous  procedure is used to identify  the profiles for the case of $(LL)_1-(HL)_2$ phase, $(LL)_2-(LH)_1$ phase and $(LL)_1-(LH)_2$ phases. 

\begin{figure}[h]
\centering
\includegraphics[height=2.0in,width=3.4in]{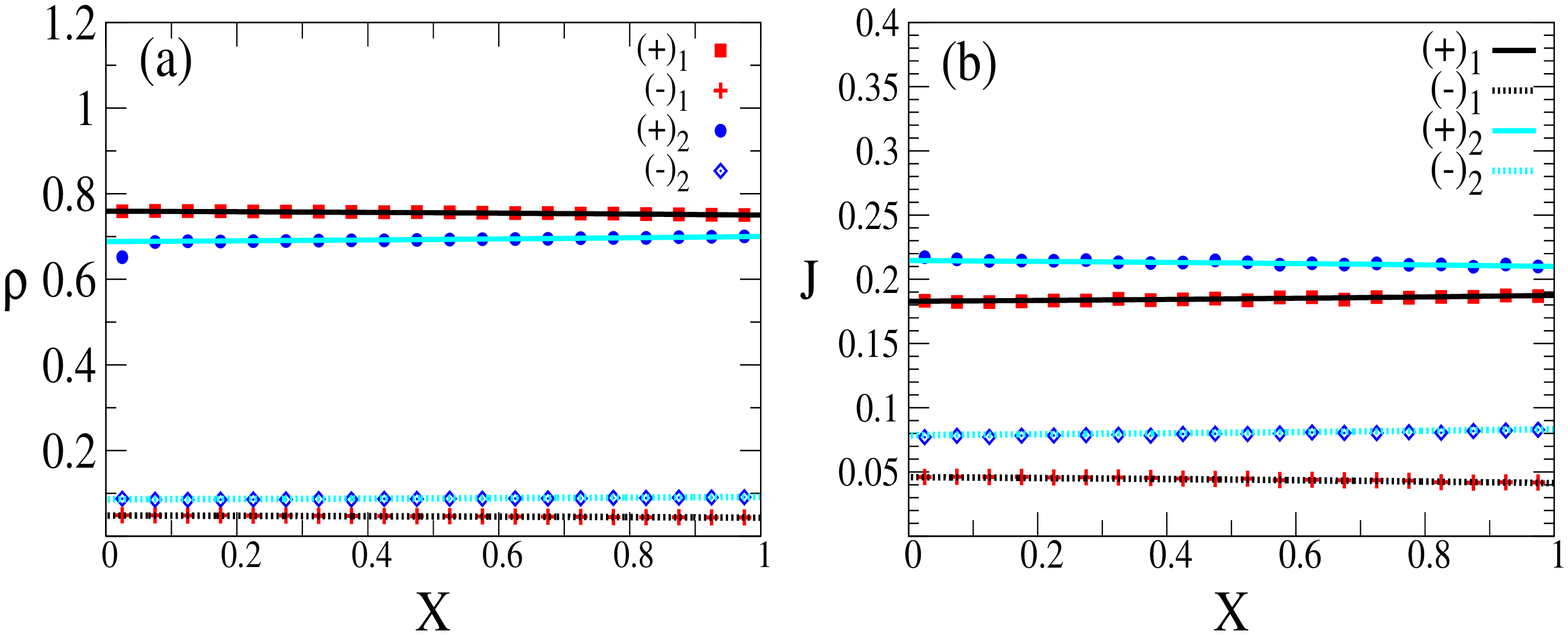}
\begin{center}
\caption{Steady state (a) density $(\rho)$ and (b) current ($J$) profile for $(+)$ and $(-)$ species in filament $1$ and $2$ as function of normalized distance $(X)$ when the system is in $(HL)_1-(HL)_2$ phase. Here, $\alpha_{1}^{+} = 0.8$, $\alpha_{1}^{-} = 0.2$, $\beta_{1}^{+} = 0.25$, $\beta_{1}^{-} = 0.7$, $\alpha_{2}^{+} = 0.9$, $\alpha_{2}^{-} = 0.4$, $\beta_{2}^{+} = 0.3$, $\beta_{2 }^{-} = 0.3$, and $\pi_{0} = 1.0$. Points are obtained by MC simulations done for system size of $N = 2000$. Solid lines are the corresponding MF solutions.}
\end{center}
\label{fig:hl_hl}
\end{figure}
\begin{figure}[h]
\centering
\includegraphics[height=2.0in,width=3.4in]{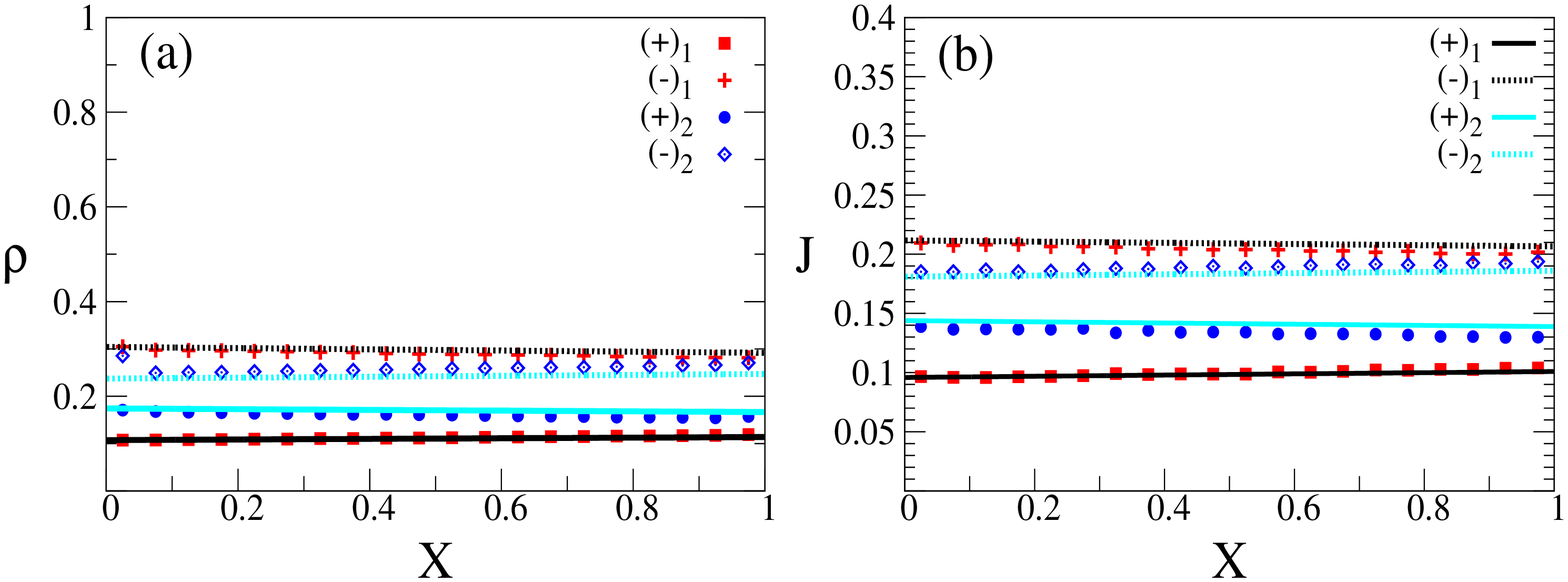}
\begin{center}
\caption{Steady state $(a)$ density and $(b)$ current profile for $(+)$ and $(-)$ when the system is in $(LL)_1-(LL)_2$ phase. Here, $\alpha_{1}^{+} = 0.2$, $\alpha_{1}^{-} = 0.4$, $\beta_{1}^{+} = 0.5$, $\beta_{1}^{-} = 0.5$, $\alpha_{2}^{+} = 0.7$, $\alpha_{2}^{-} = 0.5$, $\beta_{2}^{+} = 0.4$, $\beta_{2 }^{-} = 0.3$, and $\pi = 1.0$. Points are obtained by MC simulations done for system size of $N = 2000$. Solid lines are the corresponding MF solutions.}
\end{center}
\label{fig:ll_ll1}
\end{figure}
\begin{figure}[h]
\centering
\includegraphics[height=2.0in,width=3.4in]{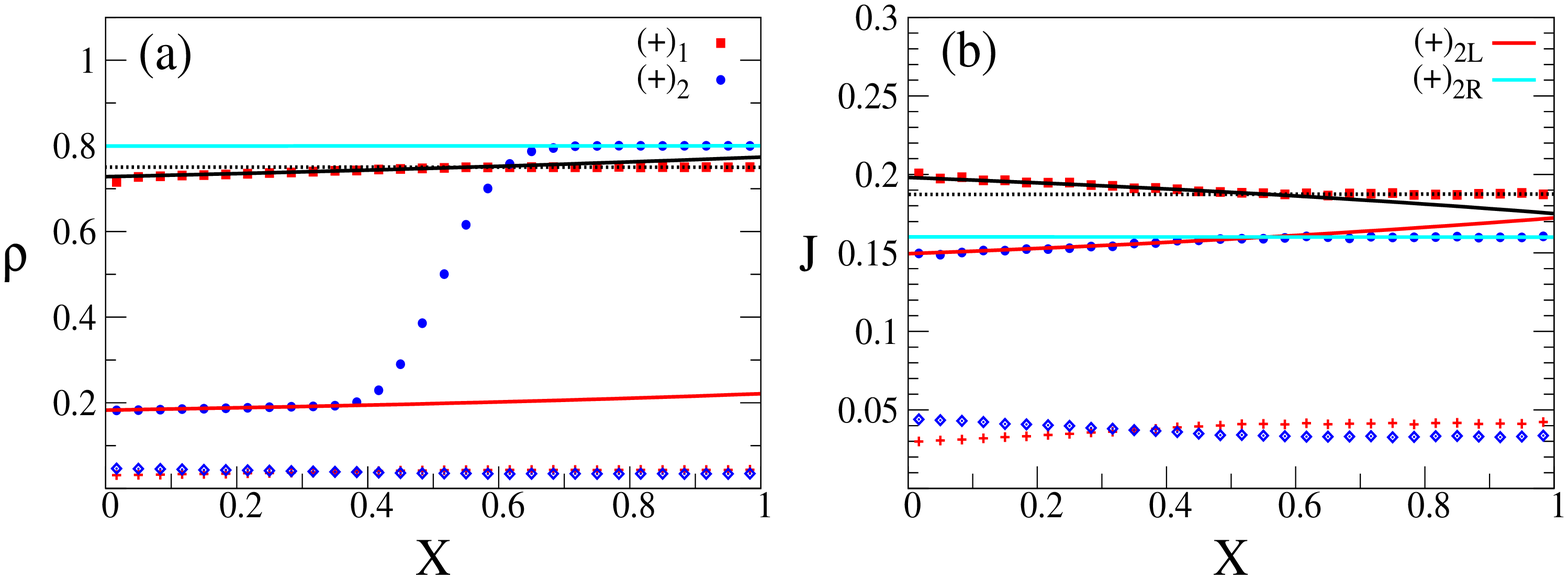}
\begin{center}
\caption{Steady state $(a)$ density and $(b)$ current profile for $(+)$ and $(-)$ when the system is in $(HL)_1-(SL)_2$ phase. Here, $\alpha_{1}^{+} = 0.45$, $\alpha_{1}^{-} = 0.2$, $\beta_{1}^{+} = 0.25$, $\beta_{1}^{-} = 0.7$, $\alpha_{2}^{+} = 0.2$, $\alpha_{2}^{-} = 0.2$, $\beta_{2}^{+} = 0.2$, $\beta_{2 }^{-} = 0.6$ and $\pi = 1.0$.Points are obtained by MC simulations done for system size of $N = 2000$. Solid lines are the MF solutions corresponding to $(HL)_1-(HL)_2$ and $(HL)_1-(LL)_2$ phases.}
\end{center}
\label{fig:hl_sl}
\end{figure}

{\it $(LL)_1-(LL)_2$ phase}: For this phase coexistence,  we need to determine the values of the boundary densities $p_{1L}$, $n_{1R}$, $p_{2L}$ and $n_{2R}$ by  solving Eq.~(\ref{eq:ASLD-LDCON3}) and use these values  to extract the associated boundary fluxes, $J_{p}$ and $J_{n}$. To this end, we assume  $n_{1L} = n_{1R}$, a symmetry that holds in the absence of particle filament exchange~\cite{evans2}. This  relation allows us to  determine $J_{1}$ and consequently fix the value of  $n_{2L}$. Again, we use the process of successive iterations for determining both $n_{1L}$ and $n_{2L}$. The entire profile can be found out by evolving densities from the left end of both filaments using Eqs.(\ref{eq:decup+1}-\ref{eq:decup-2}) (See Fig.3).

{\bf $(HL)_1-(SL)_2$}: In this phase while filament $1$ is in $HL$ phase, there is phase coexistence in the second filament, such that at the right end of the filament, the boundary condition corresponding to $HL$ phase is satisfied while at the left end of filament $2$, the boundary condition corresponding to $LL$ phase is satisfied and this phase is characterized by a density discontinuity of $(+)$ species in filament $2$ which results in a shock profile in that filament. For phase coexistence, the profile has to be such that the current for $(+)$ corresponding to HL solution equals the current for $(+)$ for the $LL$ solution at a particular spatial location between the filaments ends. This current continuity condition follows from the fact that in the bulk the total added current of the two filaments has to be conserved as there is no particle exchange with the environment. The phase coexistence can be thought of as a mixture of the 
$(HL)_1-(HL)_2$ phase with $(HL)_1-(LL)_2$. Accordingly, as one changes the parameters corresponding to input and output rate of particles, starting from a pure $(HL)_1-(LL)_2$ phase, the system can evolve into  $(HL)_1-(HL)_2$ through an intermediate phase coexistence region corresponding to the $(HL)_1-(SL)_2$ phase in  parameter space, where an incipient shock of $(+)$ particles originating at the right end of the filament eventually reaches the left end of the filament on change of parameters in the phase diagram. In order to determine the density profiles for this case, we make us of the fact that we know $p_{1R}$, $n_{1R}$, $p_{2R}$ and $n_{2R}$ because the right end of both filaments are in the  $HL$ phase. This property allows us to identify the three independent conserved currents e.g; $J_{1}$, $J_{p}$ and $J_{n}$ for the entire filaments and the density profile from the right end of the filament can be plotted using Eqs.~(\ref{eq:decup+1}-\ref{eq:decup-2}). For the left end of filament, having determined $p_{1L}$ from  Eq.~\ref{eq:ASLD-LDCON2}, the remaining  densities at the left boundary can determined using the three conserved currents. Now the  LL profile can  be simply determined starting from the left end of the filament, using Eqs.~(\ref{eq:decup+1}-\ref{eq:decup-2}). The spatial location in the bulk for which the current for this solution matches with the current for the other solution obtained for $HL-HL$ phase defines the position of shock, as shown in Fig.4. 

{\bf $(LL)_1-(SL)_2$} : In this phase while filament $1$ is in $LL$ phase, there is phase coexistence in the other filament. For filament $2$, at the right end the boundary condition corresponding to $HL$ phase is satisfied while at the left end, the boundary condition corresponding to $LL$ phase is satisfied. 
In order to determine the density profile for this case, we use Eq~.(\ref{eq:ASLD-LDCON3}) to identify  $n_{1R}$ at the right end of filament 1. Analogously, we also know $p_{2R}$ and $n_{2R}$ because filament 2 is in the $HL$ phase. At the left end of both the filaments are in $LL$ phase so that  $p_{1L}$ and $p_{2L}$ are known using Eq~.(\ref{eq:ASLD-LDCON3}). Therefore it follows that $J_{2}$, $J_{p}$ and $J_{n}$ are known for the entire filaments and all the densities at both  filament boundaries are thus determined. Now the density profiles originating from both left and right end of the filament can be plotted using Eqs.~(\ref{eq:decup+1}-\ref{eq:decup-2}) separately. The point at bulk for which the current for both these solution match defines the shock position.
\begin{figure}[h]
\centering
\includegraphics[height=2.0in,width=3.4in]{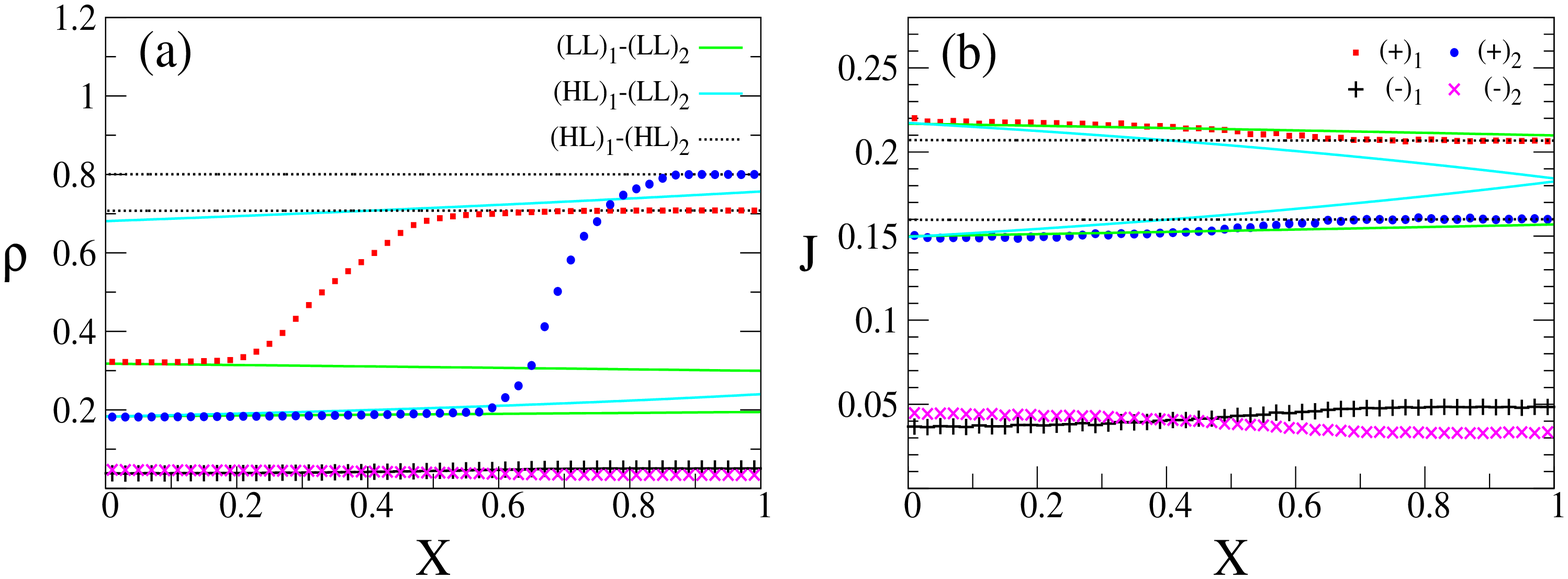}
\begin{center}
\caption{Steady state $(a)$ density and $(b)$ current profile for $(+)$ and $(-)$ when the system is in $(SL)_1-(SL)_2$ phase. Here, $\alpha_{1}^{+} = 0.35$, $\alpha_{1}^{-} = 0.2$, $\beta_{1}^{+} = 0.292$, $\beta_{1}^{-} = 0.7$, $\alpha_{2}^{+} = 0.2$, $\alpha_{2}^{-} = 0.2$, $\beta_{2}^{+} = 0.2$, $\beta_{2}^{-} = 0.6$ and $\pi = 1$. Points are obtained by MC simulations done for system size of $N = 5000$. Solid lines are the corresponding MF solutions for the phases $(HL)_1-(HL)_2$, $(HL)_1-(LL)_2$ and $(LL)_1-(LL)_2$.}
\end{center}
\label{fig:sl_sl}
\end{figure}

{\bf $(SL)_1-(SL)_2$}: This arrangement,  composed by three coexisting phases in the bulk, is characterized by shocks  for $(+)$-particles in both  filaments. In both filaments the right end  is in $(HL)_{1}-(HL)_{2}$ phase, while the region close to the left ends is a $(LL)_1-(LL)_2$ phase. In between a $(HL)_1-(LL)_2$ phase develops. The two density shocks in the bulk separate these three regions. For the phase region adjoining the right end of the filament, the  boundary condition corresponding to $(HL)_{1}-(HL)_{2}$ phase is satisfied and the boundary densities are determined by Eq.~(\ref{eq:HD1den+}) and the MF density profiles are  obtained by evolving the MF solution from the right end of the filament using these boundary densities. Accordingly,  $J_{1}$, $J_{p}$ and $J_{n}$ are known for the entire filaments. For both $(HL)_1-(LL)_2$ phase region and $(LL)_1-(LL)_2$ phase region, filament $2$ is in $LL$ phase and thus $p_{2L}$ are known using Eq.~(\ref{eq:ASLD-LDCON3}). Thus, the entire MF density and current profiles for all the species in both the filaments for the $(HL)_1-(LL)_2$ phase region can be found out using the known values of $J_{1}$, $J_{p}$ and $J_{n}$ and $p_{2L}$. For the  $(LL)_1-(LL)_2$ phase, $p_{1L}$ is known using Eq.~(\ref{eq:ASLD-LDCON3}), and along with $J_{1}$, $J_{p}$ and $J_{n}$ are used to determine the density and current profiles in this phase. The position of the shock of $(+)$ particles in filament $1$ ($x_{s1}$) is determined by matching the MF current solution of the $(HL)_{1}-(HL)_{2}$ phase with  $(LL)_{1}-(LL)_{2}$ phase at the position of the shock, 
\begin{equation}
J^{1+}_{(HL)_1-(LL)_2}(x_{s1}) = J^{1+}_{(LL)_1-(LL)_2}(x_{s1})
\end{equation}
The position of the other shock on filament $2$, $x_{s2}$, is determined by matching the MF current solution of the $(HL)_{1}-(HL)_{2}$ phase with  $(HL)_{1}-(LL)_{2}$ phase at $x_{s2}$. 
\begin{equation}
J^{2+}_{(HL)_1-(LL)_2}(x_{s2}) = J^{2+}_{(HL)_1-(HL)_2}(x_{s2})
\end{equation}
Fig.5 shows fairly good agreement for the density profiles derived from MC simulations and  the MF predictions for the $(HL)_1-(HL)_2$ and $(LL)_1-(LL)_2$. However for the $(HL)_1-(LL)_2$, the agreement between the MF solution and MC simulation does not match. 

This analysis has shown how  inter-filament switching process leads to  a wealth of new   inhomogeneous phases, allowing for phase coexistence, and the possibility of shocks in both the filaments as opposed to the collective dynamics of  transport in the absence of  such filament interactions.  

\subsection{Phase boundaries}
\label{sec:phasediagram}

In the previous subsection we have illustrated that any pair of {\it pure} phases are mediated by a phase coexistence region in the phase diagram. Whenever the current solutions corresponding to two different phases cross each other at a particular spatial location along the filament, the system exhibits phase coexistence and the spatial location of the shock coincides with the location on the filament where the two different current solutions intersect. The system selects the combination of those set of  steady state density profiles for which the corresponding current is minimum at any given spatial location in the bulk. Exploiting this insight, we can now determine the entire phase diagram and the corresponding phase boundary by using the condition when the minimum value of the current along the filaments is allowed at one of their ends. Accordingly, the phase boundary separating any two regions in the phase space corresponds to the parameter values for which the  location of the intersection of the different current solutions occurs  at either of the filament ends. We obtain the MF phase boundary using the conditions for allowed phase in particular parameter range of entry and exit rate of particles and compare these results with MC simulations.   We will concentrate on the phase diagram when varying   $(\alpha^{+}_{1}-\beta^{+}_{1})$ while holding the other parameters constant. This choice clearly illustrates the qualitative new scenarios that filament switching brings into collective transport. Moreover,  the scheme described can be straightforwardly extended to analyze the phase diagram when varying other sets of  control parameters.. 

{\em Boundary between $(HL)_1-(LL)_2$ phase with $(HL)_1-(SL)_2$}: Here the phase boundary is determined by the condition,
\begin{equation}
J^{2+}_{(HL)_1-(LL)_2}(1) =  J^{2+}_{(HL)_1-(HL)_2}(1) 
\end{equation}
where, $J^{2+}_{(HL)_1-(LL)_2}(1)$ is the current of $(+)$ species in filament $2$ at the right boundary at $x =1$, when the system is in $(HL)_1-(LL)_2$ phase and $J^{2+}_{(HL)_1-(HL)_2}(1)$ is the current of $(+)$ species in filament $2$ at $x=1$ when the system is in $(HL)_1-(HL)_2$ phase.
As discussed in the previous section,  the $(HL)_1-(SL)_2$ phase can be thought as an a mixture of the  $(HL)_1-(LL)_2$ and $(HL)_1-(HL)_2$ phases. By matching the boundary conditions for $(HL)_1-(HL)_2$, we can determine $J_p$, $J_n$ , $J_1$ and  $J^{2+}_{(HL)_1-(HL)_2}$ at $x =1$, using Eq.~(\ref{eq:LD2den-}) for the boundary densities in $(HL)_1-(HL)_2$ phase. For the $(HL)_1-(LL)_2$ phase, we use Eq.~(\ref{eq:ASLD-LDCON2}) in order to determine $p_{2L}$,  the boundary density for $(+)$ at $x=0$  in filament 2. Since there is overall particle conservation in the bulk, the values of $J_p$, $J_n$ and $J_1$ will be the same for both phases. Thus, using $p_{2L}$ along with the values of  $J_p$, $J_n$ and $J_1$, we can now find out all the boundary values for densities in both filaments at $x=0$ for the  $(HL)_1-(LL)_2$ phase. Using the evolution equations for densities, Eqs.~(\ref{eq:decup+1}-\ref{eq:decup-2}), we can find out $J^{2+}_{(HL)_1-(LL)_2}$ at $x =1$. Matching this expression for the current with $J^{2+}_{(HL)_1-(HL)_2}$  at $x =1$ determines the MF phase boundary.

{\em Boundary between $(HL)_1-(HL)_2$ phase with $(HL)_1-(SL)_2$}: Here the phase boundary is determined by the condition,
\begin{equation}
J^{2+}_{(HL)_1-(LL)_2}(0) = J^{2+}_{(HL)_1-(HL)_2}(0)
\end{equation}
where, $J^{2+}_{(HL)_1-(LL)_2}(0)$ is the current of $(+)$ species in filament $2$ at the left boundary when the system is in $(HL)_1-(LL)_2$ phase and  $J^{2+}_{(HL)_1-(HL)_2}(0)$ is the current of $(+)$ species in filament $2$ at the left boundary when the system is in the $(HL)_1-(HL)_2$ phase. The procedure for finding out the phase boundary is same as in the previous case, except that the current matching is now done at $x = 0$.

{\em Boundary between $(HL)_1-(LH)_2$ phase with $(HL)_1-(LS)_2$}: The condition for phase coexistence in this case reads 
\begin{equation}
J^{2-}_{(HL)_1-(LL)_2}(1) = J^{2-}_{(HL)_1-(LH)_2}(1),
\end{equation}
where, $J^{2-}_{(HL)_1-(LL)_2}(1)$ is the current of $(-)$ species in filament $2$ at the right boundary when the system is in $(HL)_1-(LL)_2$ phase and $J^{2-}_{(HL)_1-(LH)_2}(1)$ is the current of $(-)$ species in lane$-2$ at the right boundary when the system is in $(HL)_1-(LH)_2$ phase. For this case, $J_{1}$ and $J_{2}$ can be determined since $p_{1R}, n_{1R}, p_{2L}$ and $n_{2L}$ is known from Eq.(\ref{eq:LD2den-}). Since for $(HL)_1-(LL)_2$ phase, $n_{2R}$ is known using Eq. (\ref{eq:ASLD-LDCON2}) and $n_{1R}$ is already known using Eq.(\ref{eq:LD2den-}), thus $J_{n}$ can be determined and hence using the known values of $J_{1}$ and $J_{2}$, $J_{p}$ can also be determined. This information is sufficient to determine the density and current profiles for both $LL$ and $LH$ phase in filament 2. Therefore, the condition of current matching of $(-)$ species on filament $2$ at $x =1$ determines the location of the phase boundary.

{\em Boundary between $(HL)_1-(LL)_2$ phase with $(HL)_1-(LS)_2$}: The condition  for this phase boundary for this case is,
\begin{equation}
J^{2-}_{(HL)_1-(LL)_2}(0) = J^{2-}_{(HL)_1-(LH)_2}(0)
\end{equation}
where, $J^{2-}_{(HL)_1-(LL)_2}(0)$ is the current of $(-)$ species in filament $2$ at the left boundary when the system is in $(HL)_1-(LL)_2$ phase and $J^{2-}_{(HL)_1-(LH)_2}(0)$ is the current of $(-)$ species in filament $2$ at the left boundary when the system is in $(HL)_1-(LH)_2$ phase. The procedure to find the density and the current profiles follows exactly the arguments as in the previous case. However, the condition of current matching of $(-)$ species on filament $2$ is done at $x=0$ and this determines the phase boundary between the two phases.

{\em Boundary between $(LL)_1-(LL)_2$ phase with $(SL)_1-(LL)_2$}: The condition to determine the phase boundary is given by
\begin{equation}
J^{1+}_{(LL)_1-(LL)_2}(1) = J^{1+}_{(HL)_1-(LL)_2}(1)
\end{equation}
Here, $J^{1+}_{(LL)_1-(LL)_2}(1)$ is the current of $(+)$ species in filament $1$ at the right boundary when the system is in $(LL)_1-(LL)_2$ phase and $J^{1+}_{(HL)_1-(LL)_2}(1)$ is the current of $(+)$ species in filament $1$ at the right boundary when the system is in $(HL)_1-(LL)_2$ phase. For the $(LL)_1-(LL)_2$ phase the densities - $p_{1L}$ and $p_{2L}$  is known from Eq.~(\ref{eq:ASLD-LDCON2}), which identifies $J_p$. For  the $(HL)_1-(LL)_2$ phase $p_{1R}$,  $n_{1R}$, and $n_{2R}$ are determined by  Eq.~(\ref{eq:LD2den-}) and this is used to find $J_{1}$ and $J_n$. Having obtained these fluxes, the entire density and current profile for the two phases can be found out and the current matching condition for the $(+)$ species in filament $1$ determines the phase boundary in this case.  

{\em Boundary between $(HL)_1-(LL)_2$ phase with $(SL)_1-(LL)_2$}: Condition for this boundary in terms of boundary currents is given by
\begin{equation}
J^{1+}_{(LL)_1-(LL)_2}(0) = J^{1+}_{(HL)_1-(LL)_2}(0)
\end{equation}
Here, $J^{1+}_{(LL)_1-(LL)_2}(0)$ is the current of $(+)$ species in filament $1$ at the left boundary when the system is in $(LL)_1-(LL)_2$ phase and $J^{1+}_{(HL)_1-(LL)_2}(0)$ is the current of $(+)$ species in filament $1$ at left boundary when the system is in $(HL)_1-(LL)_2$ phase. For the $(LL)_1-(LL)_2$ phase the densities - $p_{1L}, n_{1R}, p_{2L}$ and $n_{2R}$ are known from Eq.~(\ref{eq:ASLD-LDCON2}), which identifies $J_p$ and $J_n$. For the $(HL)_1-(HL)_2$ phase, $p_{1R}$ and $n_{1R}$ are determined from  Eq.~(\ref{eq:LD2den-}) and this is used to find $J_{1}$. Having obtained these fluxes, the entire density and current profile for the two phases can be found out and the current matching condition for the $(+)$ species in filament $1$ at the left boundary determines the phase boundary.  \\\\

{\em Boundary between $(SL)_1-(LL)_2$ phase with $(SL)_1-(SL)_2$}: The phase boundary derives from the parameters for which,
\begin{equation}
J^{2+}_{(HL)_1-(LL)_2}(1) = J^{2+}_{(HL)_1-(HL)_2}(1)
\end{equation}
is satisfied,   where $J^{2+}_{(HL)_1-(LL)_2}(1)$ is the current of $(+)$ species in filament $2$ at the right boundary when the system is in the  $(HL)_1-(LL)_2$ phase and $J^{2+}_{(HL)_1-(HL)_2}(1)$ is the current of $(+)$ species in filament $2$ at the right boundary when the system is in the $(HL)_1-(HL)_2$ phase.\\
At the right boundary the system is in the $(HL)_1-(HL)_2$ phase, and therefore $p_{1R}, p_{2R}, n_{1R}$ and $n_{1R}$ can be determined using Eq.~(\ref{eq:LD2den-}). Consequently ,$J_p$, $J_n$ and $J_{1}$ are known and $J^{2+}_{(HL)_1-(HL)_2}(1)$ can be found out. Since the left end of the filament $2$ is in the $LL$ phase,  $p_{1L}$ is known, and the entire density and current profile can be determined. From these profiles we identify the remaining flux,  $J^{2+}_{(HL)_1-(LL)_2}(1)$. Matching the current for the two profiles at the right boundary determines the phase boundary.

\begin{figure}[h]
\centering
\includegraphics[height=2.0in,width=3.4in]{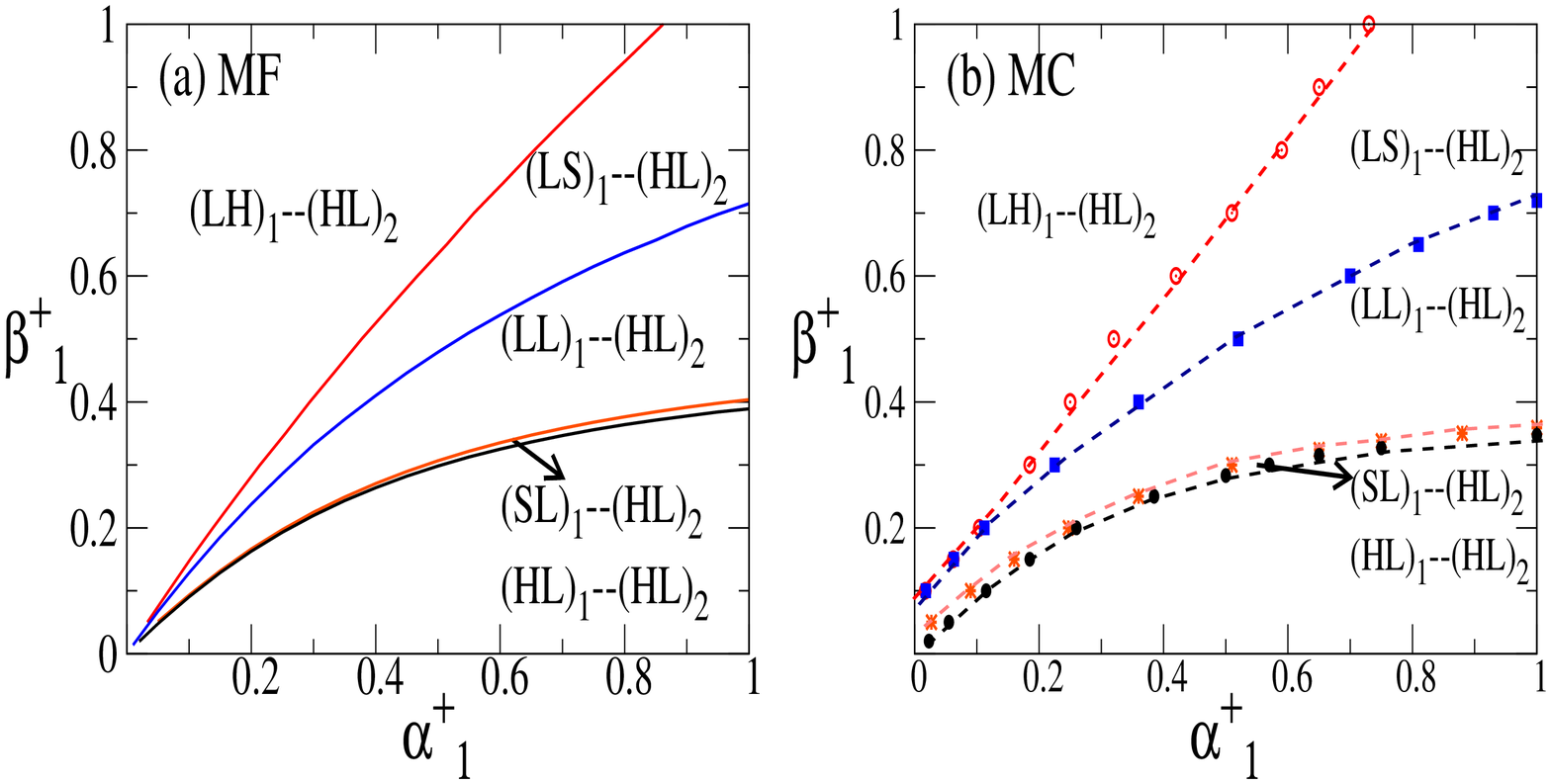}
\begin{center}
\caption{Phase space cut along $\alpha_1^+$$-$$\beta_1^+$. Here, $\alpha_{1}^{-} = 0.4$, $\beta_{1}^{-} = 0.3$, $\alpha_{2}^{+} = 0.8$, $\alpha_{2}^{-} = 0.2$, $\beta_{2}^{+} = 0.25$, $\beta_{2}^{-} = 0.7$ and $\pi = 1.0$. $(a)$ gives the $MF$ phase diagram while $(b)$ is obtained by $MC$ simulation with $N = 5000$.}
\end{center}
\label{fig:lane_1hl_mc_mf_pd}
\end{figure}
\begin{figure}[h]
\centering
\includegraphics[height=2.0in,width=3.4in]{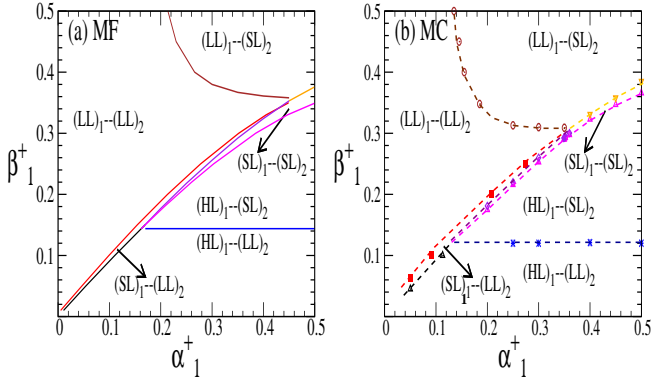}
\begin{center}
\caption{Phase space cut along $\alpha_1^+$$-$$\beta_1^+$ plane. Here, $\alpha_{1}^{-} = 0.2$, $\beta_{1}^{-} = 0.7$, $\alpha_{2}^{+} = 0.2$, $\alpha_{2}^{-} = 0.2$, $\beta_{2}^{+} = 0.2$, $\beta_{2 }^{-} = 0.6$ and $\pi = 1.0$. $(a)$ gives the $MF$ phase diagram while $(b)$ is obtained by $MC$ simulation with $N = 5000$.}
\end{center}
\label{fig:lane2_sl_mc_mf_pd}
\end{figure}

{\em Boundary between $(HL)_1-(SL)_2$ phase with $(SL)_1-(SL)_2$}: The phase boundary is derived from the condition 
\begin{equation}
J^{1+}_{(LL)_1-(LL)_2}(0) = J^{1+}_{(HL)_1-(LL)_2}(0),
\end{equation}
 where $J^{1+}_{(LL)_1-(LL)_2}(0)$ is the current of $(+)$ species in filament $1$ at the left boundary when the system is in the  $(LL)_1-(LL)_2$ phase and $J^{1+}_{(HL)_1-(LL)_2}(1)$ is the current of $(+)$ species in filament $2$ at the left boundary when the system is in the $(HL)_1-(LL)_2$ phase.\\
At the right boundary the system is in the $(HL)_1-(HL)_2$ phase. Therefore $p_{1R}$,  $p_{2R}$,  $n_{1R}$ and $n_{2R}$ can be determined using Eq.~(\ref{eq:LD2den-}), while $n_{2R}$ can be determined using Eq.~(\ref{eq:ASLD-LDCON2}), which identifies $J_p$, $J_n$ and $J_1$. Using this information the entire density and current profile can be determined for the phases $(LL)_1-(LL)_2$ and $(HL)_1-(LL)_2$. Matching the current for the two profiles in these phases at the left boundary determines the phase boundary. 

All the other possible combination of phases boundaries can  be obtained by simply interchanging the labels of filament $1$ and $2$ and using exactly the same conditions for phase boundaries that have been described in this subsection.  

From all these conditions, we can now build a complete phase diagram. Fig.6 and Fig.7 show the comparison of MF and MC phase diagram in different phase plane cuts, as a function of $(\alpha_{1}^{+}-\beta_{1}^{+})$ for fixed values of the rest of the control parameters. The MF phase boundaries that have been obtained exhibit fairly good agreement with the phase diagrams obtained by MC simulations, which shows that MF is quantitatively accurate to describe the different  phases that characterize transport intros system. The contrast between Fig.6 and Fig.7 also shows that the topology of the phase plane can drastically be altered by tuning the parameters corresponding to the particle entry and exit rates  although the filaments themselves are weakly coupled through the filament switching processes. Further, we also see that changing the particle entry and exit rates  in one filament can affect the phases in the neighbouring filament.
For instance in Fig.7 as one increases the value of entry rate of (+) particles in filament $1$, keeping the exit rate of (+) particles fixed, the resultant phase in the other filament passes over from an $LL$ phase to and $SL$ phase with a shock developing on this other filament. The phase diagram also shows the possibility to have  shock reentrant phases. For a given entry rate, the increase in the exit rate naturally  favours a transition  from  $HL$ to $LL$ phases, but  these transitions are  modulated by the developments of shocks. As a result, in the transition  from $HL$-$LL$ to $LL$-$LL$   is mediated by an intermediate region of $SL$ phases, and for large enough entry rates, the $LL$ phase is destabilized by the development of an $SL$ phase as the exit rate increases. 

\begin{figure}[h]
\centering
\includegraphics[height=2.5in,width=3.4in]{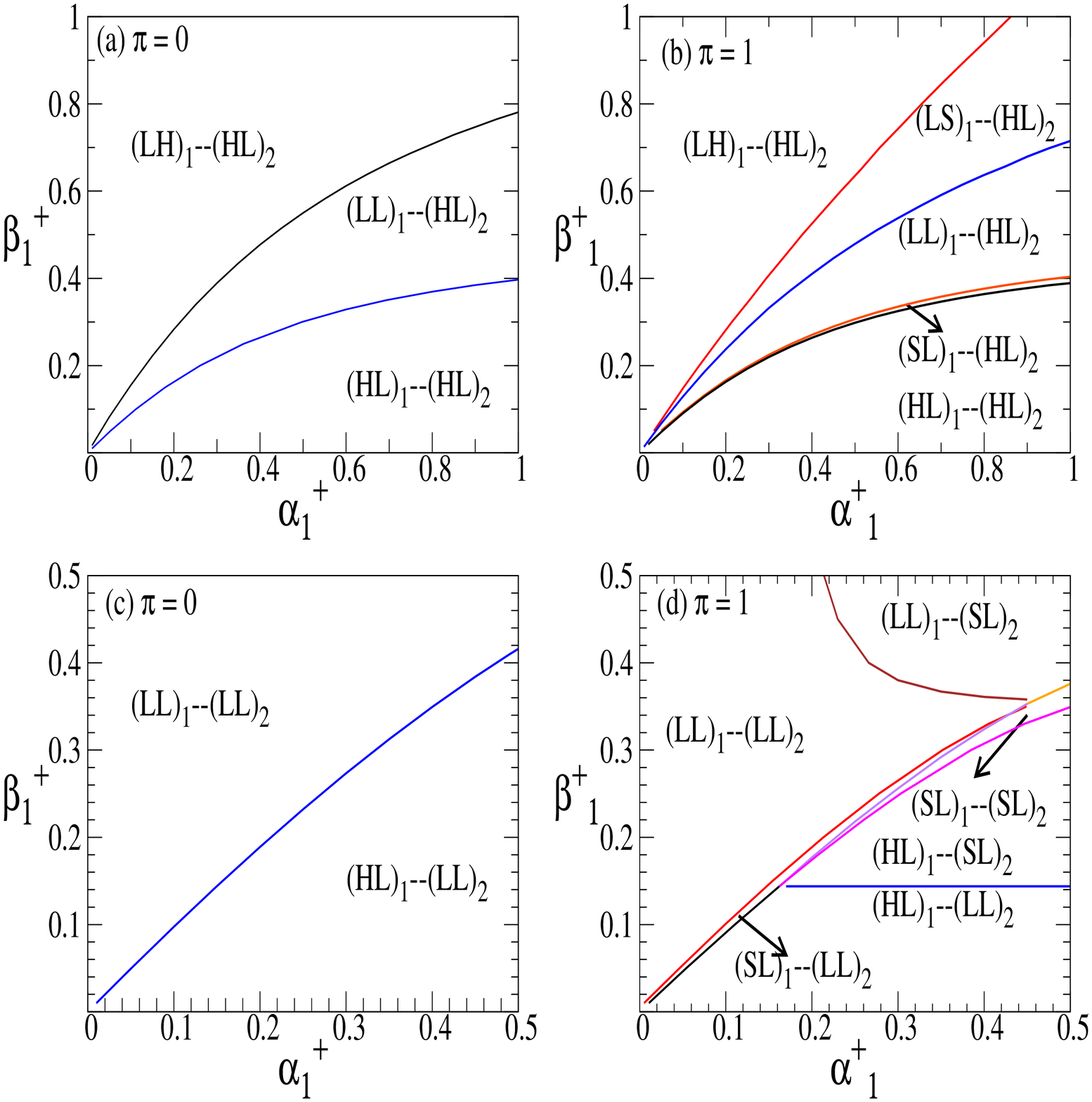}
\begin{center}
\caption{Phase plane cut along $\alpha_1^+$$-$$\beta_1^+$ plane. $[$$(a)$ and $(c)$$]$ are without filament switching $($$\pi = 0$$)$ as discussed in \cite{evans1,evans2}. [$(b)$ and $(d)$] are with filament switching $($$\pi = 1$$)$. In $(b)$; $\alpha_{1}^{-} = 0.4$, $\beta_{1}^{-} = 0.3$, $\alpha_{2}^{+} = 0.8$, $\alpha_{2}^{-} = 0.2$, $\beta_{2}^{+} = 0.25$, $\beta_{2}^{-} = 0.7$. In $(d)$;  $\alpha_{1}^{-} = 0.2$, $\beta_{1}^{-} = 0.7$, $\alpha_{2}^{+} = 0.2$, $\alpha_{2}^{-} = 0.2$, $\beta_{2}^{+} = 0.2$, $\beta_{2 }^{-} = 0.6$. Parameter regime of $(a)$ is same as that of $(c)$ and parameter regime of $(b)$ is same as that of $(d)$.}
\end{center}
\label{fig:no_ex1hl_comb}
\end{figure}
The resulting phase diagram for this system can topologically be very distinct from the phase diagrams for two species transport in the absence of particle switching between filaments.  To emphasize this fact, in Fig. 8 shows  the MF phase diagram for a system in the absence of  filament switching ($\pi=0$), Figs.8a and 8c, with corresponding  predictions when particle filament exchange is allowed  Figs.8b and 8d keep`the same weak exchange rates between the two filaments and modify the correspond to a choice of entry and exit rates. We have chosen representative parameters to show the differences associated to particle filament exchange. For Figs.8a and 8b one illustrates that one of the  new features introduced by particle exchange between filaments is the  appearance of phase coexistence  regions sandwiched between  pure phases; a scenario forbidden  when opposite particles displace along a unique filament~\cite{evans1,evans2}. Figs. 8(c) and Fig.8(d), for a different set of parameter values, show that particle filament exchange can have a deeper impact on the phase diagram topology. In this case we see that, while in the absence of filament switching dynamics the system always in a {\it pure} phase, the effect of filament switching of particles manifests as enriching the phase behaviour for the system which allows for phase coexistence and presence of shocks in the two filaments. 

As illustrated in these two examples,  generically we find that  two pure phases are  always connected by a phase region where the system exhibits phase coexistence and  bulk  localized shocks.   

\section{Conclusions}
\label{sec:conclusion}
To summarize, we have studied a multi-filament driven system with oppositely directed species of particles when the filaments are weakly coupled. Particle filament switching processes constitute correlated events because particles can only swap filaments, with a prescribed finite probability, when oppositely directed particles meet on the same filament. This aspect of filament switching mimics cellular cargo switching between neighbouring filaments during intracellular transport. We find that the interplay of the entry and exit processes of particles at the filament boundaries has a profound impact in the collective organization of  the two species of displacing particles, leading to a variety of new scenarios. Specifically, we have identified  the development  of phase coexistence on the filaments, inhomogeneous density profiles, density shocks localized in the bulk and bidirectional current flows in the system. We have developed a mean field theory (MF)  to characterize these phenomena, and have shown  that the steady state density and current profiles of particles and the phase diagram obtained using a MF formulation match reasonably well with the Monte Carlo (MC) simulation results.

While in this paper we have focused on the implications that weak coupling between filaments would have on transport when there are particle input and output in both filaments, it would be interesting to explore the regime of strong filament coupling, where we expect a weaker spatial inhomogeneity in the particle density profiles. Further, for many biological situations such as transport in axons,  it is {\it a priori} not clear whether boundary loading and off-loading of cellular cargo happens for all the parallel filaments or for specific filaments. In such situation one needs to determine the steady state distribution of cargoes and the resultant phases when one of the filament has much higher particle entry rates than the other. 

As an extreme case, we have considered a situation where one filament has closed boundaries. Starting from a random distribution of particles in both filaments, we have  observed the development of phase segregation between $(+)$ and $(-)$ particles in the closed filament. This phenomenon happens only due to the correlated lane switching process. The resulting phase segregated state in the blocked filament  does not have any flux. Following the time evolution of such a system shows that starting from a random configuration, all the vacancies are expelled and eventually the filament stops exchanging particles with the other filament as the blocked filament attains a jammed configuration, with the  $(+)$ particles piling up from the left end of the filament while the $(-)$ particles pile up on the right end . Understanding the  transition of this phase segregated jammed steady state to the steady states discussed in this paper as one slowly increases particle input and output for the filament which is initially closed at the boundaries, remains as an interesting open challenge.

\appendix

\section{Numerical simulations}
\label{sec:appendixB}
For determining the steady state density and current profiles on filaments, Monte Carlo (MC)  simulations have been performed to simulate the various processes described in Section~\ref{sec:model}. For a MC move, a filament is chosen at random and then a site in that particular filament is chosen at random with equal probability. If a particle ($+$) or ($-$) is present then a move is made for the various dynamic processes (e.g; translation or lane switching), proportional to the respective rates. We begin the simulation run starting from a random initial distribution of particles in the two filaments  and let the system evolve and reach steady state. After the system has attained steady state, averaging is done over occupation number and current in the lattice. Typically we wait for initial transient of  $1000 \frac{2 N}{q}$ swaps, where $q$ is rate of the slowest process among all the different dynamic processes occurring in the lattices. We have  further checked that the system indeed reaches its steady state by comparing the final density and current profiles at the end of the  transient. We then collect the data for occupation number and current with a period  $\geq 10 \frac{N}{q}$ and average them  over $5000$ time swaps.

In order to determine the phase boundaries by MC simulations, we use the fact  that all  phase transitions between pure phases are mediated by phase coexistence regions with shocks in the density profile. The phase boundaries can be determined numerically by tracking when the location of shock reaches the filament boundary. However, due to the finite size effects of the system, the shock has a certain finite width. We  determine the shock width and track the position of the midpoint of the shock  to identify  the phase boundaries numerically and decide when they have reached a filament end. We have used system size of $ N = 5000$ for determining the location of phase boundaries. For a fixed $\beta^+_1$ we have varied $\alpha_1^+$ in steps of $10^{-3}$ and this sets the accuracy of the phase boundaries obtained numerically.


\section{Choice of branches for the MF solution for density}
\label{sec:appendixC}
To choose a proper branch uniquely from the potential eight solutions that can be derived from   the Mean Field solutions for a particular variable we have to look carefully at the density profiles of various species in a particular phase. In Eqs.(\ref{eq:decup+1}-\ref{eq:decup-2}), notice that various combinations of $\eta^{\pm}$'s, $\mu^{\pm}$'s and $\nu^{\pm}$'s appear in this set of differential equations which govern the spatial density profile for each of the individual species. Each of the individual combinations of  $\eta^{\pm}$'s, $\mu^{\pm}$'s and $\nu^{\pm}$ appearing in these set of differential equations are exclusively functions of a single variable. For example if we consider $\eta_{p_1}^{\pm}$, then it appears in  Eq.(\ref{eq:decup+1}), where the explicit form reads as $\eta_{p_1}^{+} = \frac{1}{2} + \sqrt{\frac{1}{4} + p_1(1 - p_1) - J_1}$ and $\eta_{p_1}^{-} = \frac{1}{2} - \sqrt{\frac{1}{4} + p_1(1 - p_1) - J_1}$. Each $\eta_{p_1}^{+}$ and $\eta_{p_1}^{-}$ could separately be combined with {\em each} of the two different values of $\mu_{p_1}^{\pm}$ and $\nu_{p_1}^{\pm}$ that appear in the expressions of Eq.(\ref{eq:p1}). As there are {\em eight} possible combinations of $\eta_{p_1}^{\pm}$, $\mu_{p_1}^{\pm}$ and $\nu_{p_1}^{\pm}$, therefore  there would be eight potential solutions corresponding to the choice of a particular boundary condition. Similarly there would be eight possible solutions for Eqs.(\ref{eq:decup-1}-\ref{eq:decup-2}).     

To do the classification we have to look at the various expressions for $\eta^{\pm}$'s, $\mu^{\pm}$'s and $\nu^{\pm}$'s and they read as,
\begin{eqnarray}
\eta_{p_1}^{\pm} &=& \frac{1}{2} \pm \sqrt{\frac{1}{4} + p_1(1 - p_1) - J_1}\nonumber\\
\mu_{p_1}^{\pm} &=& \frac{1}{2} \pm \sqrt{\frac{1}{4} + p_1(1 - p_1) - J_p}\nonumber\\
\nu_{p_1}^{\pm} &=& \frac{1}{2} \pm \sqrt{(\frac{1}{2} - p_1)^2 + C_1} \label{eq:p1}
\end{eqnarray}
\begin{eqnarray}
\eta_{n_1}^{\pm} &=& \frac{1}{2} \pm \sqrt{\frac{1}{4} + n_1(1 - n_1) - J_1}\nonumber\\
\mu_{n_1}^{\pm} &=& \frac{1}{2} \pm \sqrt{\frac{1}{4} + n_1(1 - n_1) - J_n}\nonumber\\
\nu_{n_1}^{\pm} &=& \frac{1}{2} \pm \sqrt{(\frac{1}{2} - n_1)^2 - C_2} \label{eq:n1}
\end{eqnarray}
\begin{eqnarray}
\eta_{p_2}^{\pm} &=& \frac{1}{2} \pm \sqrt{\frac{1}{4} + p_2(1 - p_2) - J_2}\nonumber\\
\mu_{p_2}^{\pm} &=& \frac{1}{2} \pm \sqrt{\frac{1}{4} + p_2(1 - p_2) - J_p}\nonumber\\
\nu_{p_2}^{\pm} &=& \frac{1}{2} \pm \sqrt{(\frac{1}{2} - p_2)^2 + C_2} \label{eq:p2}
\end{eqnarray}
\begin{eqnarray}
\eta_{n_2}^{\pm} &=& \frac{1}{2} \pm \sqrt{\frac{1}{4} + n_2(1 - n_2) - J_2}\nonumber\\
\mu_{n_2}^{\pm} &=& \frac{1}{2} \pm \sqrt{\frac{1}{4} + n_2(1 - n_2) - J_n}\nonumber\\
\nu_{n_2}^{\pm} &=& \frac{1}{2} \pm \sqrt{(\frac{1}{2} - n_2)^2 - C_1} \label{eq:n2}
\end{eqnarray}

Among the  eight different branches of Eq.(\ref{eq:decup+1}) which arise due to eight possible combinations of $\eta_{p_1}^{\pm}$, $\mu_{p_1}^{\pm}$ and $\nu_{p_1}^{\pm}$, we take that particular branch of the equation which is a combination of $\eta_{p_1}^-$, $\mu_{p_1}^-$ and $\nu_{p_1}^-$. This branch would be classified as Sol$-1$. The entire nomenclature is classified in Table$-$\ref{table:branches} for different solutions of Eq.(\ref{eq:decup+1}). The same nomenclature is true for the different solutions of Eqs.(\ref{eq:decup-1}-(\ref{eq:decup-2}).

\begin{table}[ht]
\caption{Classification of the branches} 
\centering 
\begin{tabular}{|c|c|c|c|}
\hline\hline 
$\eta$ & $\mu$ & $\nu$ & Branch \\ [0.5ex] 
\hline 
$\eta_{p_1}^{-}$ & $\mu_{p_1}^{-}$ & $\nu_{p_1}^{-}$ & Sol$-1$ \\ 
\hline 
$\eta_{p_1}^{-}$ & $\mu_{p_1}^{-}$ & $\nu_{p_1}^{+}$ & Sol$-2$ \\ 
\hline 
$\eta_{p_1}^{-}$ & $\mu_{p_1}^{+}$ & $\nu_{p_1}^{-}$ & Sol$-3$ \\ 
\hline 
$\eta_{p_1}^{+}$ & $\mu_{p_1}^{-}$ & $\nu_{p_1}^{-}$ & Sol$-4$ \\ 
\hline 
$\eta_{p_1}^{-}$ & $\mu_{p_1}^{+}$ & $\nu_{p_1}^{+}$ & Sol$-5$ \\ 
\hline 
$\eta_{p_1}^{+}$ & $\mu_{p_1}^{-}$ & $\nu_{p_1}^{+}$ & Sol$-6$ \\ 
\hline 
$\eta_{p_1}^{+}$ & $\mu_{p_1}^{+}$ & $\nu_{p_1}^{-}$ & Sol$-7$ \\ 
\hline 
$\eta_{p_1}^{+}$ & $\mu_{p_1}^{+}$ & $\nu_{p_1}^{+}$ & Sol$-8$ \\
 [1ex] 
\hline 
\end{tabular}
\label{table:branches} 
\end{table}

 Let us assume that filament $1$ is in LL phase whereas filament $2$ is in HL phase. Therefore the density of $(+)$ and $(-)$ particles in both the filaments and hence the variables will have the following values.
\begin{eqnarray}
p_1 < \frac{1}{2}, n_1 < \frac{1}{2}\nonumber\\
p_2 > \frac{1}{2}, n_2 < \frac{1}{2} \nonumber
\end{eqnarray}
Thus, the proper choice of roots, in this case, while substituting other variables in terms of $p_1$ would be
\begin{eqnarray}
n_1 &=& \frac{1}{2} - \sqrt{\frac{1}{4} + p_1(1 - p_1) - J_1} \equiv \eta_{p_1}^{-}\nonumber\\
p_2 &=& \frac{1}{2} + \sqrt{\frac{1}{4} + p_1(1 - p_1) - J_p} \equiv \mu_{p_1}^{+}\nonumber\\
n_2 &=& \frac{1}{2} - \sqrt{(\frac{1}{2} - p_1)^2 + C_1} \equiv \nu_{p_1}^{-}\\\nonumber
\end{eqnarray}
Thus sol$-3$ is the correct branch for $p_1$ when the system is in $(LL)_1-(HL)_2$ phase. Following the same procedure we can show that in this particular phase sol$-2$, sol$-1$ and sol$-4$ are the proper choices for the variables $n_1$, $p_2$ and $n_2$ respectively. Relevant branches for all other phases are depicted in Table$-$\ref{table:phase_sol}.\\

\begin{table}[ht]

\caption{MF Solutions In Different Phases} 
\centering
\begin{tabular}{|c|c|c|c|c|}
\hline\hline 
Phase & $p_1$ & $n_1$ & $p_2$  & $n_2$\\ [0.5ex] 
\hline 
$(LL)_1-(LL)_2$ & Sol$-1$ & Sol$-1$ & Sol$-1$ & Sol$-1$ \\
\hline
$(LL)_1-(HL)_2$ & Sol$-3$ & Sol$-2$ & Sol$-1$ & Sol$-4$ \\
\hline
$(LL)_1-(LH)_2$ & Sol$-2$ & Sol$-3$ & Sol$-4$ & Sol$-1$ \\
\hline
$(HL)_1-(HL)_2$ & Sol$-3$ & Sol$-6$ & Sol$-3$ & Sol$-6$ \\
\hline
$(HL)_1-(LH)_2$ & Sol$-2$ & Sol$-7$ & Sol$-7$ & Sol$-2$ \\
\hline
$(LH)_1-(LH)_2$ & Sol$-6$ & Sol$-3$ & Sol$-6$ & Sol$-3$ \\
[1ex] 
\hline 
\end{tabular}
\label{table:phase_sol} 
\end{table}

\newpage

{\bf Acknowledgments}\\
SM acknowledges DBT RGYI Project No: BT/PR6715/GBD/27/463/2012 for financial support. IP acknowledges financial support from
MINECO (Spain), Project FIS2011- 22603, DURSI Project 2009SGR-634, and Generalitat
de Catalunya under program Icrea Academia.

\end{document}